# Effects of temperature and surface orientation on migration behaviours of helium atoms near tungsten surfaces

Xiaoshuang Wang, Zhangwen Wu, Qing Hou[*]

*Key Lab for Radiation Physics and Technology, Institute of Nuclear Science and Technology,*

*Sichuan University, Chengdu 610064*

**ABSTRACT**

Molecular dynamics simulations were performed to study the dependence of migration behaviours of single helium atoms near tungsten surfaces on the surface orientation and temperature. For W{100} and W{110} surfaces, He atoms can quickly escape out near the surface without accumulation even at a temperature of 400 K. The behaviours of helium atoms can be well-described by the theory of continuous diffusion of particles in a semi-infinite medium. For a W{111} surface, the situation is complex. Different types of trap mutations occur within the neighbouring region of the W{111} surface. The trap mutations hinder the escape of He atoms, resulting in their accumulation. The probability of a He atom escaping into vacuum from a trap mutation depends on the type of the trap mutation, and the occurrence probabilities of the different types of trap mutations are dependent on the temperature. This finding suggests that the escape rate of He atoms on the W{111} surface does not show a monotonic dependence on temperature. For instance, the escape rate at $T=1500$ K is

---

[*] Corresponding author. Tel.: +86 28 85412104; fax: +86 28 85410252.

   *E-mail address*: qhou@scu.edu.cn (Q. Hou)

lower than the rate at $T=1100$ K. Our results are useful for understanding the structural evolution and He release on tungsten surfaces and for designing models in other simulation methods beyond molecular dynamics.

# 1. Introduction

The interaction of helium with tungsten surfaces is an issue that has draw intensive attention in the research and development of nuclear fusion reactors because tungsten is a candidate for plasma-facing material and will undergo a high flux irradiation with H and He. Many studies have shown that the exposure of tungsten surfaces to helium plasma induces various morphological changes on the surfaces, depending on the exposure conditions [1-5]. These morphological changes could give rise to changes in thermal and mechanical properties of the surfaces, as well as the production of impurities and dusts that would be a potential contamination source of fusion plasma.

The morphological evolution of the surfaces upon irradiation with helium is essentially a multi-scale phenomenon involving a large number of various atomistic processes that interplay with each other [6]. To achieve a comprehensive understanding of the morphological evolution of the surfaces, detailed descriptions on each of the fundamental atomistic processes are thus critical. Because direct experimental observation of the atomistic processes is difficult, multi-scale computer simulations play an important role. As an indispensable tool in the chain of multi-scale simulations, molecular dynamics (MD) simulations can give a detailed view of atomic mechanisms for the morphological evolution of surfaces and also can provide the dynamic parameters for simulation methods such as the kinetic Monte Carlo (KMC) or the rate theory (RT), which can account for the evolution of surfaces over a longer time period.

There have been a number of MD simulations that deal with the interaction of helium atoms with tungsten [7-18]. The He atoms, starting from their incidence on a surface, are generally considered to experience two major phases. The first phase is the transport process in which the incident energetic He atoms lose their kinetic energies due to collisions with the substrate atoms. Some of the He atoms are propelled out of the substrate with a certain kinetic energy, and the others are retained in the substrate with the kinetic energy of the surrounding atoms. The transport process takes approximately 2 picoseconds. The reflection and depth distribution for low energy (<200eV) He bombardments on W surfaces have been studied by Li et al. [7], Borovikov et al. [13] and Hammond et al. [11] using MD simulations. The retained He atoms were found to be distributed at a depth of a few nms. The depth distributions obtained by Li et al. were shallower than those obtained by the latter studies probably due to the energy exchange between the He projectiles and electrons that had been included in the MD simulations of Li et al. The second phase is the diffusion process in which the retained He atoms, which are driven by temperature, move randomly in the substrate until it escapes out of the surfaces or becomes trapped in sites such as vacancies or helium clusters. There is also chance for a helium atom to dissociate from a trapping site and to return to diffusion status [15]. For the diffusion behaviours of the He atoms and the small He clusters in the bulk W, MD studies were conducted by Wang et al[19], Zhou et al[15] and Perez et al[16]. These studies found the non-Arrhenius diffusion of He atoms at high substrate temperatures and some counter-intuition features for the migration of small He clusters (e. g., $He_5$ cluster migrates faster than $He_4$ cluster).

The purpose of the present paper is to study the near surface behaviour of He atoms in the second phase. In the study of Li et al. for the reflection and depth distribution of He atoms on W {100} surfaces [7], an accumulation of He atoms at a depth of 3~4 monolayers below the surfaces was observed. This observation was explained as follows: a He projectile that backscattered from deeper locations may induce a replacement sequence in the <111> direction with the consequence that the He atom may substitute a W atom and a W atom stacks on the surface [8]. In this paper, we focus on whether a diffusing interstitial atom, which could be the He projectiles that have been slowed down to the environmental temperature or the He atoms dissociated from the trapping sites as mentioned above, could also accumulate on the surfaces. The MD study of Hu et al. [10] has shown that small He clusters migrating toward the W surfaces may dissociate and form He-vacancy clusters near the surface and induce stacking W atoms. Soon after, similar dynamic process was observed by Barashev *et al*. [12], using self-evolving atomistic kinetic Monto Carlo (SEAKMC) method. Hammond and Wirth [11] also found accumulations of He near the W surfaces and stacking of the W atoms on the surfaces. They mentioned that the stacking of W atoms can be induced even by single He atoms migrating to the surfaces. However, the parameter space they addressed was very limited because only three temperatures were considered. To bridge the MD simulations with other simulation methods, such as KMC or RT, a more quantitative description of the near surface behaviour of He atoms is needed.

In the present paper, MD simulations were performed for the migration of single

He atoms near W surfaces of different orientations and temperatures. The states of the He atoms in the substrates were identified by their centro-symmetry parameters, and the probabilities for the He atoms getting trapped near the surfaces were extracted. The rate dependence for the He atoms to escape out of the substrates on the surface orientation and temperature was analysed and quantified based on the theory for continuous diffusion and thermal desorption.

## 2. Simulation Methods

All the MD simulations were performed using the MDPSCU, a molecular dynamics package that can run on multiple graphic processing units (GPUs) for parallel computing [20]. The Finnis-Sinclair type potential proposed by Ackland and Thetford [21], which has been widely used in the literature [10, 12, 13, 15-17], was adopted for the interactions between W atoms. For the interactions between the He and W atoms, we used a pairwise potential that was constructed by fitting its long range part to *ab initio* data and smoothly connecting the long range part to Ziegler–Biersack–Littmark (ZBL) potential in the short range part [19].

The simulation procedure contains two major phases. In the first phase, the simulation boxes were generated for bcc W with three box sizes: $10a_0 \times 10a_0 \times 30a_0$, $11.31a_0 \times 10a_0 \times 28.28a_0$ and $11.31a_0 \times 9.79a_0 \times 31.18a_0$ adopted respectively, where $a_0 = 3.1652 \overset{\circ}{\text{A}}$ is the lattice constant of tungsten. Corresponding to the three box sizes, the z-axis of the boxes, which would be the normal direction of the surfaces in the second step, is defined in the (100), (110) and (111) crystal orientations. A box side in the z-direction was set longer than in x- and y-direction to eliminate possible interactions between the surfaces that would appear in the second step. For each of the box sizes, we generated 1000 independent replicas. In every box, one He atom was introduced at a position randomly selected in the box. The simulation boxes were thermalized to a given temperature ranging from 200 K to 2500 K at first, and then relaxed for enough time steps to bring them to thermal equilibrium. Thermalisation of a simulation box was conducted by assigning the atoms in the box with velocities generated by the Monte Carlo sampling of the Maxwell distribution of atom velocity. In the present phases, periodic boundary conditions were applied in all the x-, y- and z-directions. After the simulation boxes were prepared in the first phase, the second phase started with the periodic boundary condition in the z-direction discarded. Thus, every simulation box in the second phase contained two free surfaces in the opposite z-direction, and the He atoms had a uniform initial depth distribution that was convenient for analysing the simulation results as shown below. The constructed boxes evolved for 500 ps using a time step of 1 fs. The states of the boxes were recorded for every 5ps.

## 3. Results and Discussion

Based on the methods described above, MD simulations were performed. We are interested in the effects of temperature and surface orientation on the migration of helium atoms. In the following subsections, we first discuss the migration of He atoms near the {100} and {110} surfaces.

**3.1.** Migration of He atoms near the W{100} and W{110} surfaces

Fig. 1a. displays two snapshots that were generated by merging snapshots of 1000 simulation boxes of {100} surfaces at the end of the simulations (500 ps) for temperatures of $T$=400 K and 1000 K, respectively. Although a simulation box has two surfaces, only the left surface of the boxes was drawn. Fig. 1b. is the merged snapshot for the W{110} surfaces at the same time and temperature. We observed that no He atom stays in the first two crystal layers of the surface. This phenomena was also observed in our simulations for other substrate temperatures. There is only one box with the W{100} surface and at $T$=1000 K in which a substitutional He atom is found in the third monolayer below the surface along with a stacking W atom found in the <111> direction from the He atom. More quantitatively, Fig .2 displays a snapshot of the depth distributions of He in the simulation boxes for $T$=1000 K at the end of the simulations ($t$=500 ps). The depth distributions for the W{111} surface will be discussed later. It should be noted that the coordinate origin is the centre of the boxes. The depth distributions for the W{100} and W{110} surfaces exhibit no accumulation of He atom near the surfaces, also suggesting that the He atoms can escape out of the substrate in < 5ps once they migrate to the second monolayer of the {100} and {110} surfaces, and thus, the formation of pairs of substitutional He and stacking W atoms is a rare event. This observation is consistent with the findings in Ref. [11] and [10] in which He clusters, not single He atoms, were found to induce pairs of substitutional He and stacking W atoms on the W{100} or W{110} surface with a reasonable probability. In contrast, in the study by Li *et al.,* for bombardments of single He atoms

on W {100} surfaces [8], the accumulation of He atoms at a depth of 3~4 layers below the surfaces was found. The difference between what was observed in the work of Li *et al.* and what is observed in the present paper is due to that a He atom during the transport process, which was simulated in the former work, could have a residual kinetic energy higher than the threshold for generating a replacement sequence along the <111> direction. However, in this work, a thermally diffusing interstitial He atom hardly generates such a sequence near the W{100} and W{110} surfaces and the He atoms remain interstitial.

Because He atoms can quickly escape out of the substrates as described above when they migrate to the second or third monolayer under the surfaces, the effect of the surfaces can be modelled by an absorbing layer. The He atoms can be considered to be migrating with a diffusion coefficient, *D,* in the bulk tungsten until arriving at the absorbing layer and are absorbed instantly with the probability denoted by *R*. Based on the continuous diffusion theory [22], we deduced the number of absorbed He atoms, $N_a$, at time *t*:

$$N_a(t) = N_B \left[ 1 - \int_{-h}^{h} dz \frac{n_0(z)}{2} (E_1 + E_2 + E_3 + E_4) \right] \quad (1)$$

where

$$E_1 = \sum_{k=0}^{\infty} (1-2R)^k \, erf\left[ \frac{h+(-1)^k(2kh-z)}{\sqrt{4Dt}} \right]$$

$$E_2 = \sum_{k=0}^{\infty} (1-2R)^k \, erf\left[ \frac{h+(-1)^{k+1}(2kh-z)}{\sqrt{4Dt}} \right]$$

$$E_3 = \sum_{k=1}^{\infty} (1-2R)^k \, erf\left[ \frac{h+(-1)^{k+1}(2kh+z)}{\sqrt{4Dt}} \right]$$

$$E_4 = \sum_{k=1}^{\infty}(1-2R)^k \, erf\left[\frac{h+(-1)^k(2kh+z)}{\sqrt{4Dt}}\right].$$

The $N_B$ is the number of simulation boxes. In the present paper, $N_B =1000$. The $n_0(z)$ is the initial depth distribution of the He atoms. Because the He atoms were initially introduced into the simulation boxes at randomly selected positions as described in Section 2, the $n_0(z)=1/L_z$ with $L_z$ as the simulation box size in the z-direction. The $h$ is the distance from the centre of the simulation box to the absorbing layer. Assuming the thickness of the absorbing layer is small in comparison with $L_z$, we defined $h=L_z/2$. For more details on the deduction of eq. (1), one may refer to the Appendix.

The absorbing probability, $R$, for {100} and {110} surfaces was extracted by fitting eq. (1) to the MD simulation data and is summarized in Table 1 for the different substrate temperatures. Because the He atoms escape almost instantly out of the substrates through the absorbing layer, we used $N_e(t)$, the number of He atoms out of the substrate, instead of $N_a(t)$. For the diffusion coefficient $D$ in bulk, we used the value given by Zhou *et al.*[15]. Considering that there are He atoms in the first two layers and the near surface W atoms require time to relax to equilibrium at the beginning of the second phase of simulations (refer to section 2), we thus took $t=5$ ps as the starting time of the diffusion-escape process of He atoms and discarded the He atoms escaping out of the substrate before this time. Moreover, considering that the diffusion behaviour of near surface He atom may deviate from their diffusion behaviour in bulk, only the simulation data of $t>200$ ps, which were mainly contributed from He atoms at deeper locations, were adopted in the fitting process.

Fig. 3 shows the comparison between the number of escaping He atoms as a function of time obtained by eq. (1) vs. the MD simulations. The analytical results are found in good agreement with the simulation results for $t>200$ ps. For small $t$ ($<200$ ps), the MD results are generally smaller than what is predicted by eq. (1) and the $R$ value, shown in Table 1, is smaller than 1 for both W{100} and W{110} surfaces. In the viewpoint of continue diffusion theory, the value $R$ suggests that the absorbing layer is not a perfect absorber and there is a probability for a He atom being reflected back into bulk by the surface. The inserts in Fig. 3 depict the potential of the system as a function of the depth of He atom below the W{100} and W{110} surfaces. We obtained the potential-depth relations by MD simulations, in which the system was relaxed at zero temperature with W atoms allowed to move in three dimensions and the He atoms allowed to move only in x- and y- directions at a number of fixed z-positions. It is observed that the potential barrier closest to the surface is higher than the barriers at deeper locations. This gives the reason for the absorbing layer deviating from a perfect absorbing boundary in continue diffusion theory in which the probability for an atom jumping forward and backward is the same. Grossly, the value of $R$ increases with increasing temperature, indicating the surface effect tends to diminish with an increasing substrate temperature.

In summary, the interstitial He atoms migrate with the diffusion coefficient in bulk until they migrate to the second monolayer below the W{100} and W{110} surfaces. The effect of the first two layers of the surfaces can be modelled by an absorbing boundary of continue diffusion theory, which absorbs the He atoms on the boundary with a probability smaller than 1. Once a He atom is absorbed by the absorbing layer, the He atom almost instantly escape into vacuum. This is informative to establish models in KMC and the boundary condition in the rate theory. For an example in KMC, the time for a He atom staying in the first two layers below the W{100} and W{110} surfaces can be excluded from the characteristic time calculation of the occurrence of an event[23]. The escape of an interstitial He atom occurs with probability *R* when the selected atom jumps into the neighbouring region of the W{100} and W{110} surfaces (i.e., two monolayers below the surfaces). However, the situation for the W{111} surface is different where trap mutations play a role.

**3.2.** Migration of He atoms near the W{111} surface

The effect of the W{111} surface is also like an absorbing layer to the in-substrate He atoms. The He atoms are absorbed by the absorbing layer and then escape into vacuum. However, unlike what occurs in the W{100} and W{110} surfaces, most of the He atoms cannot instantly escape into vacuum through the absorbing layer. Instead, the He atoms may accumulate in the absorbing layer as shown in Fig. 2c for *T*=1000 K and *t*=500 ps. The hindered escaping of the He atoms, which is dependent on the temperature, is due to the formation of pairs of substitutional He and stacking (or interstitial) W atoms near the W{111} surface. For the convenience of description, we also use the term "trap mutation" [10, 24] for the formation of pairs of substitutional and stacking (interstitial) atoms. Fig. 4 shows the typical trap mutations found in our simulations, denoted by $S_n$, $M_n$ and $L_n$, respectively, with the subscript *n* denoting the trap mutations occurring on the *n*th monolayer under the surface. The stacking W atoms in all types of trap mutations occupy the stacking position of the S layers. We used the centrosymmetry parameter (CSP), which was first introduced by Kelchner *et al.* to determine the local structures in the fcc crystals [25], to identify the state of the neighbouring region of He atoms. It had been shown by Moriarty *et al.* that dislocations in void growth can be more clearly indentified by CSP than by atomic energy characterization [26]. In the present work, we also observed that the $S_1$ trap mutations and $M_2$ can be distinguished by their CSP values but not by their potentials.

According to the definition of CSP, CSP for a He atom in a bcc W is written as:

$$CSP = \sum_{j=1}^{4}|\mathbf{R}_j + \mathbf{R}_{-j}|^2 + \sum_{k=1}^{3}|\mathbf{R}_k + \mathbf{R}_{-k}|^2 \qquad (2)$$

where *j* denotes the four pairs of the first bcc neighbouring sites and *k* denotes the three pairs of the second bcc neighbouring sites of the He atom. The $\mathbf{R}_j$ is the vector from site *j* to the nearest atom (a W atom in the present paper) of this site. To diminish the uncertainties caused by thermal fluctuation, we quenched the simulation boxes to zero temperature before calculating the CSP. In the temperature range from 200 K to 1400 K, the CSP is 0.1~0.3 for $S_1$, $2.5\times10^{-4}$~$1\times10^{-3}$ for $S_2$ and $1\times10^{-6}$~$1\times10^{-4}$ for $M_3$ trap mutations, respectively. For $M_2$ trap mutations, CSP may vary between 0.72~1.6 and $1\times10^{-3}$~$2.24\times10^{-3}$, corresponding to the transformation from a pair of interstitial He and interstitial W atoms to a pair of substitutional He and stacking W atoms. Such a situation was also observed for the $L_2$ trap mutations with CSP varying between 0.6~0.72 and $2.24\times10^{-3}$~$4\times10^{-2}$. For an interstitial He atom in the bulk, its CSP is typically in the range of 3.1~3.7. For temperatures higher than 1400 K, it is difficult to identify the types of trap mutations even if the configurations have been quenched to zero temperature. However, we can also identify a near surface He atom if its CSP <3. We also distinguished the He atoms according to their escaping time. A near surface He atom is defined as quick-escaping if the He atom escapes out of the substrate in the time interval of recording, otherwise the He atom is a long-staying atom. Almost all He atoms in trap mutations are long-staying atoms.

Based on the above criteria, the absorbed He atoms by the absorbing layer were counted. Fig. 5 demonstrates the number $N_a(t)$ of absorbed He atoms for the W{111} surface as a function of time obtained by the MD simulations. The $N_a(t)$~t relation for the W{111} surface is similar to that for the W{100} and W{110} surfaces. The MD simulation data in Fig. 5 were fitted also using eq. (1) in the same fitting process described above for W{100} and W{110} surfaces, and the values of $R$ defined in eq. (1) are listed in Table 2. The $R$ value for the W{111} surface is slightly larger than 1 at a temperature greater than 400 K. A larger $R$ value of 1.342 was found at $T$=400 K. The insert in Fig. 5 shows the system potential as a function of the depth of He atom below the W{111} surface. In contrast to what have been shown in Fig.3, the potential barriers near the surface (denoted by A, B, C) are slightly lower than the barriers at deeper depths. The feature of the potential-depth relation gives a qualitative explanation for $R$>1. Even so, the results indicate that the absorbing layer model is also validated for W{111}. However, the He atoms absorbed by the W{100} and W{110} absorbing layers escape out of the surface instantly, while most of the absorbed He atoms in the W{111} absorbing layer are trapped in the absorbing layer for a longer time.

The fractions of the trap mutations and quick-escaping He occurring near the W{111} surface as a function of temperature in the range of 400 K~1400 K were calculated and are displayed in Table 3 and Fig. 6a. Generally, the fraction of quick-escaping He is small in this temperature range. At low temperatures ($T$<1000 K), the $S_1$ trap mutation is the dominant event occurring near the W{111} surface, while very few He atoms can quickly escape out of the substrates. With an increasing temperature, trap mutations other than the $S_1$ type appear, and the fraction of $S_1$ trap mutations decrease rapidly for $T$>1000 K. Above $T$=1350 K, more $M_2$ trap mutations are observed than $S_1$ trap mutations. The $M_2$ trap mutations hinder the escaping of the He atoms out of the substrates more than the $S_1$ trap mutations. Furthermore, Fig .6b displays the ratio, $R_e$, of He atoms escaping into vacuum over the absorbed He atoms (including quick-escaping He atoms) also at a time of 500 ps as a function of temperature. The dependence of $R_e$ on $T$ is in contrast to the intuition idea that $R_e$ would increase monotonically with increasing $T$. The $R_e$ is very small for $T$<600 K because few He atoms in $S_1$ trap mutations can escape at this temperature. The $R_e$ increases significantly starting from $T$=800 K and reaches a maximum value around $T$=1100 K. For $T$>1100 K, more than 80% of the He atoms trapped in the $S_1$ trap mutations escape into vacuum during the simulation period (500 ps). After $T$=1100 K, $R_e$ decreases to a minimum value around $T$=1500 K due to the decreasing percentage of $S_1$ trap mutations and the increasing percentage of $M_2$ mutations. He atoms do not escape from the M$_2$ trap mutations until $T$=1400 K, and at this time, a small fraction (9.3%) of He atoms in M$_2$ trap mutations are found to escape into vacuum. Above $T$=1500 K, $R_e$ exhibits nearly a linear increase with increasing temperature, a behaviour that is similar to Brownian diffusion.

The above results indicate that He atoms may stay for a long time near the W{111} surface. If the time that the He atoms are near the surface is comparable with the production rate of the near surface He atoms, it is likely that He clusters will form. In the KMC simulations, the trap mutations and escaping He atoms should be included in the event sampling. To extract the rate, which is denoted by $R_s$ with the subscript representing the type of trap mutation, for the He atoms in the trap mutations that escape into vacuum, we propose a MD simulation method that simulates the thermal desorption experiments. In this method, the temperature of the simulation boxes was changed according to $T=T_0+\beta t$ by scaling the velocities of the atoms in the boxes, where $T_0$ is the initial temperature and $\beta$ is the parameter controlling the rate of temperature change. We assume that $R_s$ follows Arrhenius behaviour as follows: $R_s = A_s \exp(-E_s^{(a)}/k_B T)$, and the equation

$$2\ln T_m - \ln \beta = E_s^{(a)}/k_B T_m + \ln(E_s^{(a)}/k_B A_s) \qquad (3)$$

can be written according to the theory of thermal desorption, where $T_m$ is the temperature at which the thermal desorption spectrum has a maximum value. Using the quenched simulation boxes mentioned above in which the S$_1$ trap mutations were identified, we extracted the activation energy, $E_{s1}^{(a)}$, and the prefactor, $A_{s1}$, for the trap mutation, S$_1$, on the basis of eq. (3) with a set of $\beta$ varying from 2.43 to 8.50 K/ps.

Fig. 7a shows the accumulated He atoms escaping out of the substrates as a function of temperature obtained by our MD simulation of the thermal desorption of He atoms in the $S_1$ trap mutations. Using Fig. 7a, the $T_m(\beta)$ was determined for different $\beta$ values and Fig. 7b was drawn according to eq. (3). Although the value $\beta$ in our MD simulations is much larger than the rate of temperature increase in any thermal desorption experiments, we observe that eq. (3) is fitted well to the simulation data with $E_{s1}^{(a)}$=0.605 eV and $A_{s1}$=4.941 ps$^{-1}$. Using the same method with $\beta$ varying from 2.84 to 4.5 K/ps, we also calculated the activation energy, $E_{M2}^{(a)}$, and the prefactor, $A_{M2}$, for the trap mutation, $M_2$. Again, eq. (3) is fitted well to the simulation data (Fig. 7c) with $E_{M2}^{(a)}$=1.589 eV and $A_{M2}$= 43.238 ps$^{-1}$.

## 4. Conclusions

We performed MD simulations that demonstrate the dependence of migration behaviour of single near surface He atoms on the surface orientation and temperature of W. To the in-substrate He atoms, a surface can be modelled by an absorbing layer that absorbs He atoms when the He atoms migrate to the surface in 3~4 monolayers. The absorbing rates can be fit well using the continuous diffusion theory. However, the states of the absorbed He atoms strongly depend on the surface orientations. For the W{100} and {110} surfaces, almost all the absorbed He atoms escape out of the substrate instantly, while the situation is much more complex for the W{111} surface. For the W{111} surface, depending on temperature, trap mutations most likely occur with only a small fraction of absorbed He atoms quickly escaping into vacuum through the absorbing layer. The occurrence probabilities of different types of trap mutations depend on temperature. For a temperature of T<1000K, the $S_1$ trap mutation is dominant. With increasing temperature, the fraction of $S_1$ trap mutation diminishes and other types of trap mutations appear, which may hinder the escaping of He atoms more. The change in the fraction of different types of trap mutations may lead to an escaping rate of He atoms on the W{111} surface that does not have a monotonic dependence on temperature. For instance, the escaping rate at $T$=1500 K is smaller than the rate at $T$=1100 K. These results are useful for understanding the structural evolution and He release of tungsten surfaces, especially in the initial phase, and for designing models in the methods of time-space scales beyond what the MD simulations can achieve. However, to have a comprehensive quantitative description of the structural evolution and He release of tungsten surfaces, more MD simulation data are required, such as for other surface orientations.

**Acknowledgments**

This work was supported by the National Magnetic Confinement Fusion Program of China (2013GB109002).

**Appendix**

Based on the random walk model, the problems of particles diffusing in infinite and semi-infinite medium have been discussed in detail by S. Chandrasekhar [22]. For infinite medium, if there is a point particle source of unit intensity at position $z_0$ and $t=0$, the particle density distribution after time $t$ is given by

$$W(z,t;z_0) = \frac{1}{2(\pi Dt)^{1/2}} \exp\left(-(z-z_0)^2/4Dt\right) \tag{A.1}$$

where $D$ is the diffusion coefficient. For semi-infinite medium, the effect of the boundary of the semi-infinite medium can be modeled by a wall, which is located at $z=h$ and denoted by B1 (Fig.A.1), that would reflect or absorb a particle whenever the particle arrives at the wall. According to S. Chandrasekhar, $W(z, t)$ can be rewritten as:

$$W(z,t;z_0,h) = \frac{1}{2(\pi Dt)^{1/2}} \left\{ \exp\left(-(z-z_0)^2/4Dt\right) \pm \exp\left[-\left(z-(2h-z_0)\right)^2/4Dt\right] \right\} \tag{A.2}$$

The plus and minus sign of the second term in eq. (A.2) correspond to reflection and absorption at the wall, respectively.

We noted that the positions $2h-z_0$ and $z_0$ are mirror symmetric relative to the wall. Thus, the effect of a wall on the density distribution in the medium is equivalent to a virtual point particle source that is placed at $z=2h-z_0$ and $t=0$ in an infinite medium. The intensity of the virtual particle source is positive if the wall is a perfect reflection wall, and the intensity is negative if the wall is a perfect absorption wall. Assuming that the wall can absorb only a fraction (denoted by $R$) of the particles that arrive at the boundary, we can rewrite eq. (A.2) in a more general form:

$$W(z,t;z_0 h) = \frac{1}{2(\pi Dt)^{1/2}} \left\{ \exp\left(-(z-z_0)^2/4Dt\right) + (1-2R)\exp\left[-\left(z-(2h-z_0)\right)^2/4Dt\right] \right\} \tag{A.3}$$

The intensity of the virtual particle source becomes (*1-2R*) time of the intensity of the original source.

Now, let us consider the case in which there are two boundaries, B1 and B2, located at *z=h* and *z=-h*, respectively. Based on the virtual source model given above, the original particle source at position $z_0$ would have two virtual sources of intensity (*1-2R*) at *z=2h-z₀* and *z=-2h-z₀*. The virtual source at *z=2h-z₀* would also have a virtual source of intensity (*1-2R*)² at *z=-2h-(2h-z₀)=-4h+z₀* in the mirror symmetrical to B2. Likewise, the virtual source at *z=-4h+z₀* would have an virtual source of intensity (*1-2R*)³ at *z=2h-(-4h+z₀)=6h- z₀* in the mirror symmetrical to B1. Similarly, we can continue the process further. After applying the same process to the virtual source at *z=-2h-z₀*, an infinite number of virtual sources can be finally obtained. The positions of the virtual sources are expressed as $(-1)^{k+1}(\pm 2kh - z_0)$ and the intensities are $(1-2R)^k$ with the number *k* running from 1 to infinite. With the original particle source and the virtual sources, we can write the particle density distribution for a medium containing two boundaries:

$$W(z,t;z_0,h) = \frac{1}{2(\pi Dt)^{1/2}}\left\{\exp\left(-(z-z_0)^2/4Dt\right)+C_1+C_2\right\} \quad for \quad -h \leq z \leq h \qquad (A.4),$$

where:

$$C_1 = \sum_{k=1}^{\infty}(1-2R)^k \exp\left[-\left(z-(-1)^{k+1}(2kh-z_0)\right)^2/4Dt\right],$$

$$C_2 = \sum_{k=1}^{\infty}(1-2R)^k \exp\left[-\left(z-(-1)^{k+1}(-2kh-z_0)\right)^2/4Dt\right].$$

If the initial particle source is not a point source but has a normalized distribution,

$n(z_0)$, the particle density distribution is then:

$$W(z,t;h) = \int_{-h}^{h} dz_0 \, n(z_0) W(z,t;h,z_0) \quad for \quad -h \leq z \leq h \qquad (A.5).$$

After some simplification, we can deduce the normalized number of particles remaining in the medium as:

$$W(t;h) = \int_{-h}^{h} dz \int_{-h}^{h} dz_0 \, n(z_0) W(z,t;h,z_0) = \int_{-h}^{h} dz_0 \frac{n(z_0)}{2} \{E_1 + E_2 + E_3 + E_4\} \qquad (A.6)$$

where,

$$E_1 = \sum_{k=0}^{\infty} (1-2R)^k \, erf\left[\frac{h+(-1)^k (2kh-z)}{\sqrt{4Dt}}\right]$$

$$E_2 = \sum_{k=0}^{\infty} (1-2R)^k \, erf\left[\frac{h+(-1)^{k+1} (2kh-z)}{\sqrt{4Dt}}\right]$$

$$E_3 = \sum_{k=1}^{\infty} (1-2R)^k \, erf\left[\frac{h+(-1)^{k+1} (2kh+z)}{\sqrt{4Dt}}\right]$$

$$E_4 = \sum_{k=1}^{\infty} (1-2R)^k \, erf\left[\frac{h+(-1)^k (2kh+z)}{\sqrt{4Dt}}\right]$$

**Reference**

[1] S. Takamura, N. Ohno, D. Nishijima, and S. Kajita, Plasma Fusion Res. 1 (2006) 051.
[2] M. J. Baldwin and R. P. Doerner, Nucl. Fusion 48 (2008) 035001.
[3] M. J. Baldwin, R. P. Doerner, D. Nishijima, K. Tokunaga, and Y. Ueda, J. Nucl. Mater. 390-391 (2009) 886.
[4] S. J. Zenobia and G. L. Kulcinski, Phys. Scr. T138 (2009) 014049.
[5] S. Kajita, W. Sakaguchi, N. Ohno, N. Yoshida, and T. Saeki, Nucl. Fusion 49 (2009) 095005.
[6] S. J. Zinkle, Phys. Plasmas 12 (2005) 058101.
[7] M. Li, J. Wang, and Q. Hou, J. Nucl. Mater. 423 (2012) 22.
[8] M. Li, J. Cui, J. Wang, and Q. Hou, J. Nucl. Mater. 433 (2013) 17.
[9] L. Hu, K. D. Hammond, B. D. Wirth, and D. Maroudas, J. Appl. Phys. 115 (2014) 173512.
[10] L. Hu, K. D. Hammond, B. D. Wirth, and D. Maroudas, Surf. Sci. 626 (2014) L21.
[11] K. D. Hammond and B. D. Wirth, J. Appl. Phys. 116 (2014) 143301.
[12] A. V. Barashev, H. Xu, and R. E. Stoller, J. Nucl. Mater. 454 (2014) 421.
[13] V. Borovikov, A. F. Voter, and X.-Z. Tang, J. Nucl. Mater. 447 (2014 ) 254.
[14] J. Wang, Y. Zhou, M. Li, and Q. Hou, J. Nucl. Mater. 427 (2012) 290.
[15] Y. L. Zhou, J. Wang, Q. Hou, and A. H. Deng, J. Nucl. Mater. 446 (2014) 49.
[16] D. Perez, T. Vogel, and B. P. Uberuaga, Phys. Rev. B 90 (2014) 014102.
[17] F. Sefta, N. Juslin, and B. D. Wirth, J. Appl. Phys. 114 (2013) 243518.
[18] F. Sefta, K. D. Hammond, N. Juslin, and B. D. Wirth, Nucl. Fusion 53 (2013) 073015.
[19] J. Wang, Y. Zhou, M. Li, and Q. Hou, J. Nucl. Mater. 427 (2012) 290.
[20] Q. Hou, M. Li, Y. Zhou, J. Cui, Z. Cui, and J. Wang, Comput. Phys. Commun. 184 (2013) 2091.
[21] G. J. Ackland and R. Thetford, Philos. Mag. A 56 (1987) 15.
[22] S.Chandrasekhar, Rev. Mod. Phys. 15 (1943) 1.
[23] C. S. Becquart and C. Domain, J. Nucl. Mater. 385 (2009) 223.
[24] K. O. Henriksson, K. Nordlund, and J. Keinonen, Nucl. Instrum. Methods B 244 (2006) 377.
[25] C. L. Kelchner, S. J. Plimpton, and J. C. Hamilton, Phys. Rev. B 58 (1998) 11085.
[26] J. F. B. John A Moriarty, Robert E Rudd, Per Soderlind, Frederick H Streitz and Lin H Yang, J. Phys.: Condens. Matter 14 (2002) 2825.

**Captions**

**Table 1**. The R values obtained for the fits using eq. (1). The diffusion coefficient used in eq. (1) is given in ref. [15].

**Table 2.** The R values obtained for the fits using eq. (1) for the W{111} surfaces. The diffusion coefficient used in eq. (1) is the same as in Table 1.

**Table 3**. The percentage of trap mutations and quick-escaping occurring near the W{111} surface.

**Fig. 1**. Merged snapshots of 1000 simulation boxes at the end of the simulation time for the temperature of $T$=400 K and 1000 K, respectively. Dark dots: W atoms; Dark circles: He atoms. (a) W{100} surface; (b) W{110} surface.

**Fig. 2.** Histogram of the depth distribution of helium atoms for $T$=1000 K at $t$=500 ps for the (a) W {110}; (b)W{110}; and (c) W{111} surfaces.

**Fig. 3.** The accumulated number of He atoms escaping out of the substrate as a function of time. Symbols are the data obtained by the MD simulations. □:400K; ○:600K; △:800K; ▽:1000K; ◇:1200K. Solid lines were obtained by fitting eq. (1) to the MD data. (a) for the W{100} surface; (b) for the W{110} surface.

**Fig. 4.** Schematics of the trap mutations occurring near the W{111} surface observed in the MD simulations. Hollow circles: W atoms on their lattice positions; dark filled circles: He atoms; grey

solid circles: displaced W atoms; grey solid circles without border: the original positions of the displaced W atoms; and arrows: denote moving direction of displaced W atoms. The grey solid circles without border sometimes are covered by grey solid circles. Star symbols denote the higher CSP of $M_2$ and $L_2$.

**Fig. 5.** The accumulated number of He atoms absorbed by the absorbing layer of the W{111} surface as a function of time for different temperatures. □:400K; ○:600K; △:800K; ▽:1000K; ◇:1200K. Solid lines were obtained by fitting eq. (1) to the MD data.

**Fig. 6.** (a) The percentage of different types of trap mutations and quick-escaping He atoms of the total absorbed He atoms as a function of temperature; (b) The percentage of He atoms escaping into vacuum of all the absorbed He atoms.

**Fig. 7.** (a) The accumulated number of He atoms escaping into vacuum from $S_1$ trap mutations in MD simulations of thermal desorption. The arrows point to the $T_m$.; (b) $2\ln T_m - \ln\beta$ vs. $1/T_m$ for the $S_1$ trap mutations and (c) for the $M_2$ trap mutations. Symbols: the data of MD simulations; Solid line: eq. (3) fitted to the MD data (refer to eq. (3) in the text).

**Fig. A.1**. A schematic diagram explaining the concept of virtual source.

**Table 1**. The R values obtained for the fits using eq. (1). The diffusion coefficient used in eq. (1) is given in ref. [15].

| Temperature(K) | Diffusion coefficient[*] (cm$^2$/s) | R for W{100} | R for W{110} |
|---|---|---|---|
| **400** | $1.17 \times 10^{-5}$ | 0.801 | 0.754 |
| **600** | $4.33 \times 10^{-5}$ | 0.881 | 0.846 |
| **800** | $7.16 \times 10^{-5}$ | 0.934 | 0.945 |
| **1000** | $9.34 \times 10^{-5}$ | 0.826 | 0.907 |
| **1200** | $1.07 \times 10^{-4}$ | 0.878 | 0.937 |

**Table 2.** The R values obtained for the fits using eq. (1) for the W{111} surfaces. The diffusion coefficient used in eq. (1) is the same as in Table 1.

| Temperature (K) | Diffusion coefficient* (cm$^2$/s) | R for W{111} |
|---|---|---|
| **400** | $1.17 \times 10^{-5}$ | 1.342 |
| **600** | $4.33 \times 10^{-5}$ | 1.029 |
| **800** | $7.16 \times 10^{-5}$ | 1.049 |
| **1000** | $9.34 \times 10^{-5}$ | 1.078 |
| **1200** | $1.07 \times 10^{-4}$ | 1.105 |

**Table 3**. The percentage of trap mutations and quick-escaping occurring near the W{111} surface.

| Temperature (K) | Reaction types (%) | | | | | |
| --- | --- | --- | --- | --- | --- | --- |
| | Quick escaping | S1 | M2 | L2 | S2 | M3 |
| **200** | -- | 100 | -- | -- | -- | -- |
| **400** | 0.97 | 99.03 | -- | -- | -- | -- |
| **600** | 0.68 | 99.32 | -- | -- | -- | -- |
| **800** | 0.83 | 95.59 | 2.20 | 0.28 | 1.10 | -- |
| **850** | 2.01 | 92.48 | 2.51 | 0.25 | 2.51 | 0.25 |
| **900** | 5.12 | 86.45 | 5.12 | 0.26 | 3.07 | -- |
| **950** | 4.72 | 82.13 | 7.44 | 0.25 | 4.72 | 0.74 |
| **1000** | 7.03 | 75.88 | 11.94 | 0.94 | 3.28 | 0.94 |
| **1100** | 12.68 | 57.42 | 18.90 | 2.87 | 5.02 | 3.11 |
| **1200** | 18.03 | 37.53 | 28.51 | 4.61 | 5.03 | 6.29 |
| **1350** | 23.67 | 18.15 | 32.94 | 10.45 | 4.93 | 9.86 |
| **1400** | 14.50 | 19.75 | 31.72 | 15.55 | 7.98 | 10.50 |

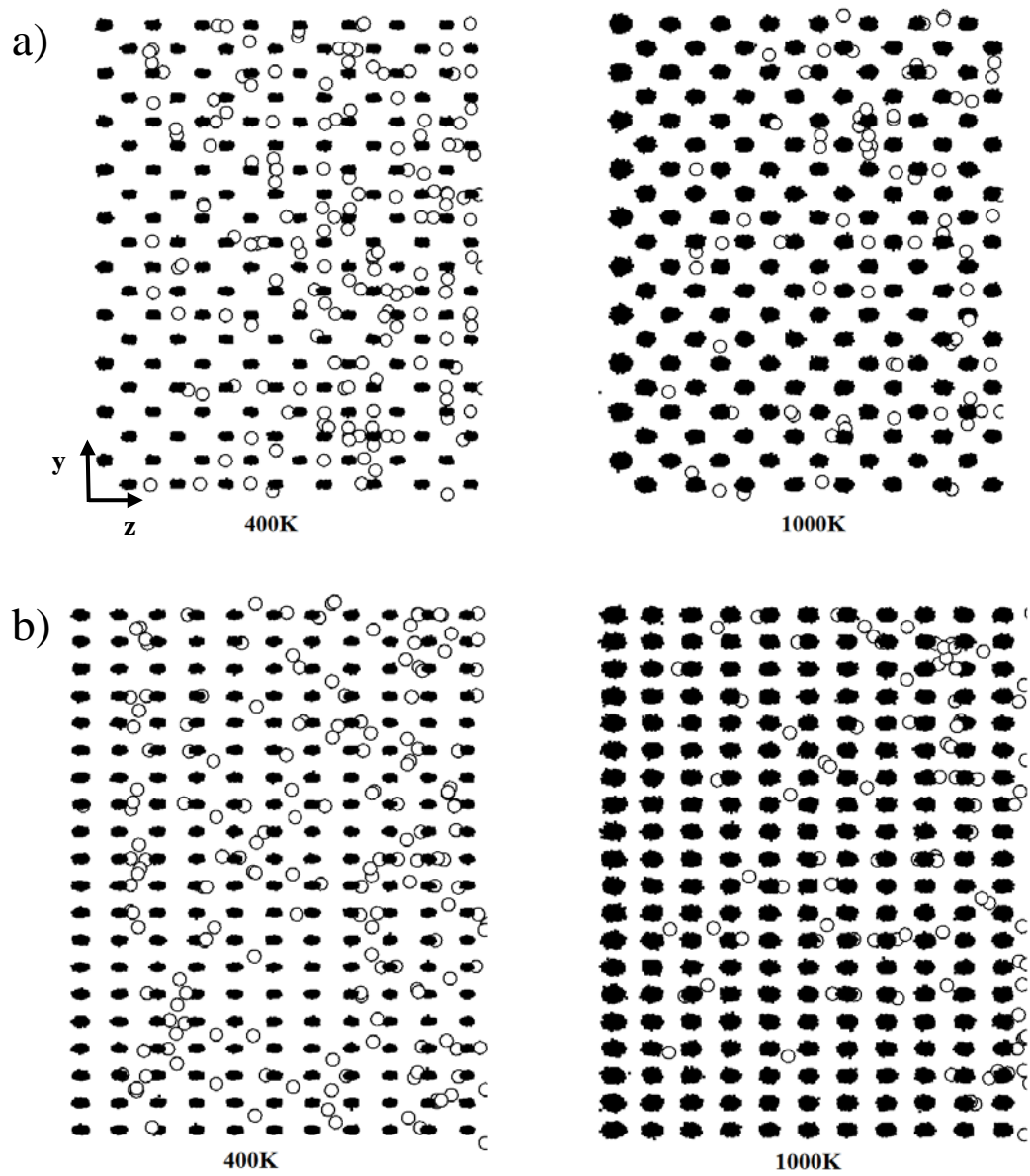

**Fig. 1**. Merged snapshots of 1000 simulation boxes at the end of the simulation time for the temperature of $T$=400 K and 1000 K, respectively. Dark dots: W atoms; Dark circles: He atoms. (a) W{100} surface; (b) W{110} surface.

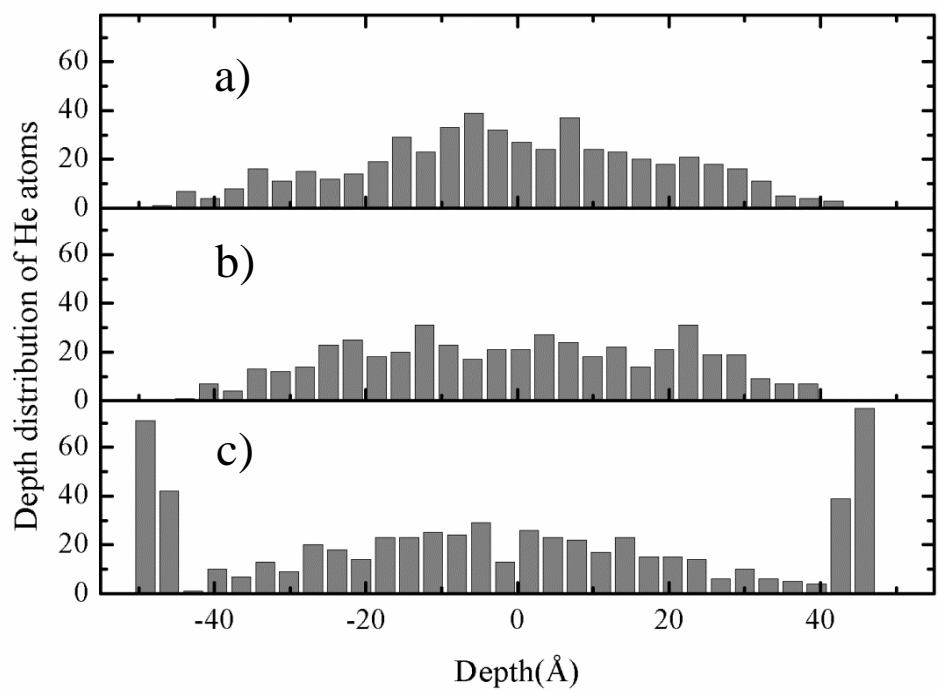

**Fig. 2.** Histogram of the depth distribution of helium atoms for *T*=1000 K at *t*=500 ps for the (a) W {110}; (b)W{110}; and (c) W{111} surfaces.

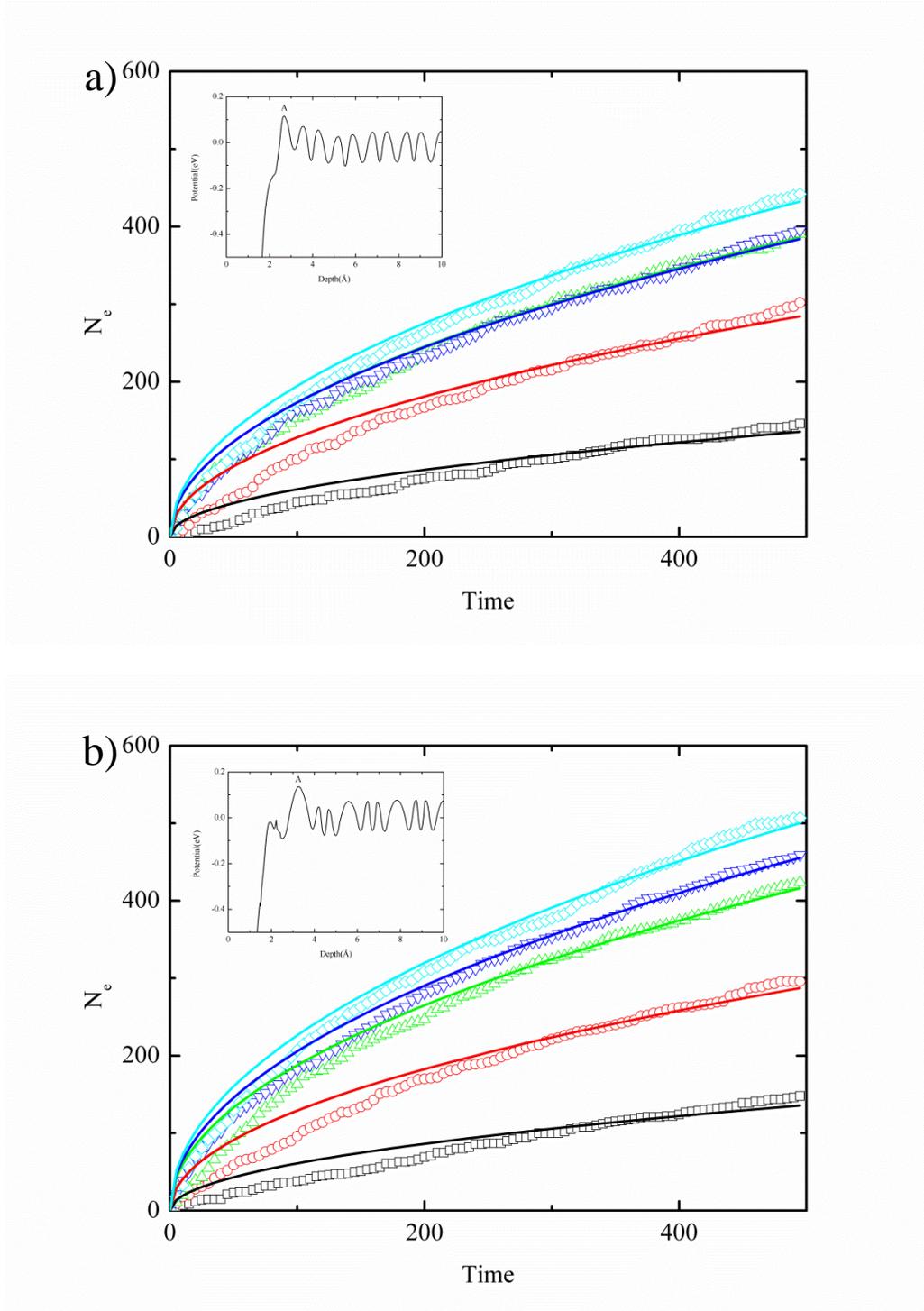

**Fig. 3.** The accumulated number of He atoms escaping out of the substrate as a function of time. Symbols are the data obtained by the MD simulations. □:400K; ○:600K; △:800K; ▽:1000K; ◇:1200K. Solid lines were obtained by fitting eq. (1) to the MD data. (a) for the W{100} surface; (b) for the W{110} surface.

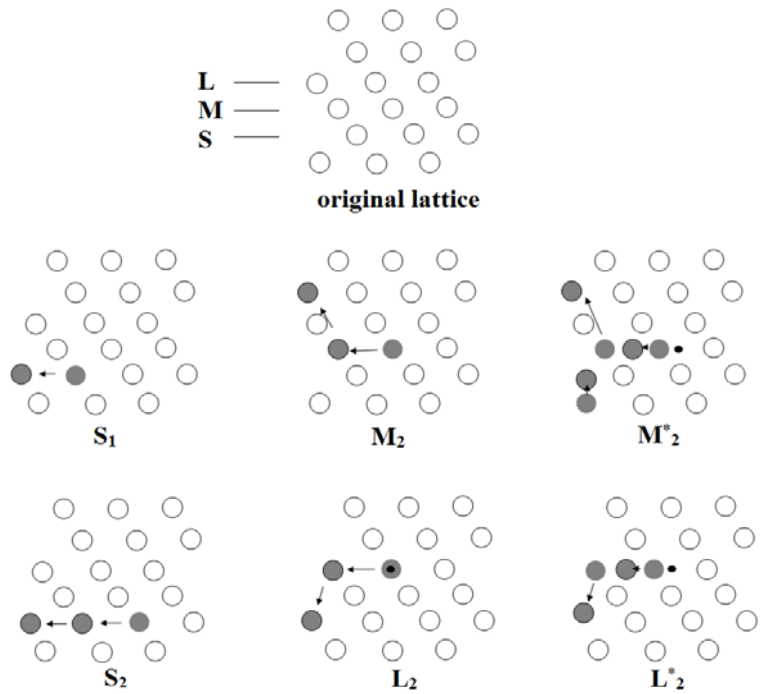

**Fig. 4.** Schematics of the trap mutations occurring near the W{111} surface observed in the MD simulations. Hollow circles: W atoms on their lattice positions; dark filled circles: He atoms; grey solid circles: displaced W atoms; grey solid circles without border: the original positions of the displaced W atoms; and arrows: denote moving direction of displaced W atoms. The grey solid circles without border sometimes are covered by grey solid circles. Star symbols denote the higher CSP of $M_2$ and $L_2$.

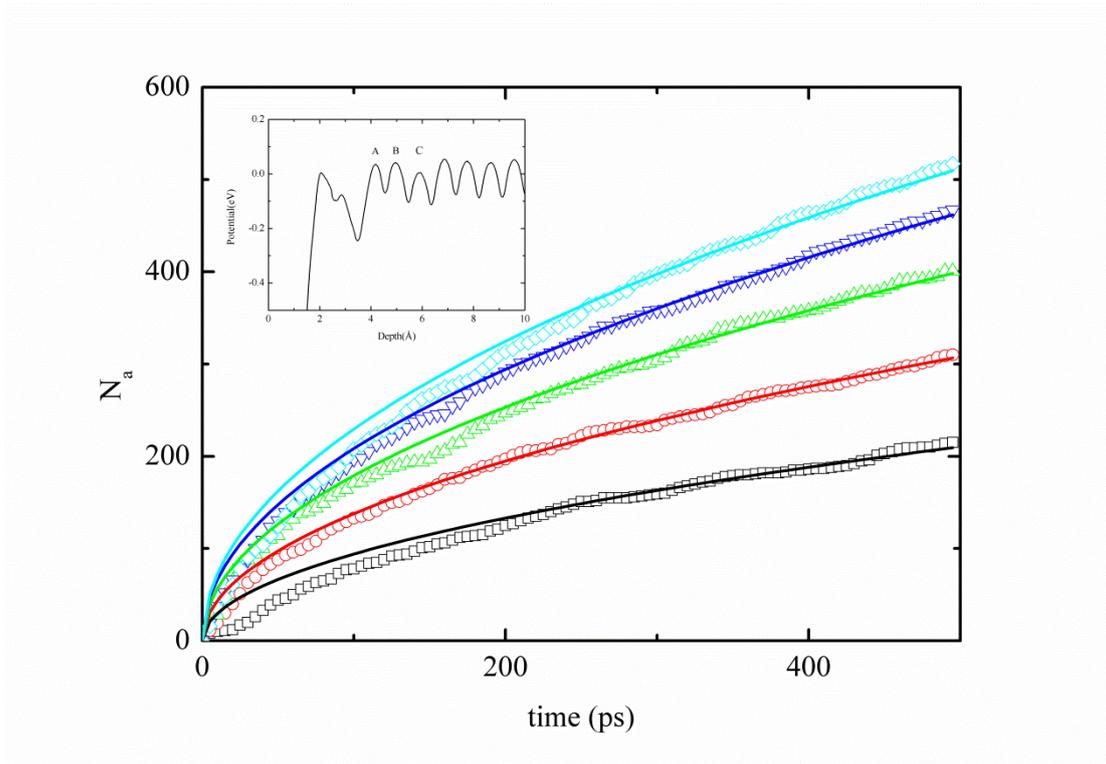

**Fig. 5.** The accumulated number of He atoms absorbed by the absorbing layer of the W{111} surface as a function of time for different temperatures. □:400K; ○:600K; △:800K; ▽:1000K; ◇:1200K. Solid lines were obtained by fitting eq. (1) to the MD data.

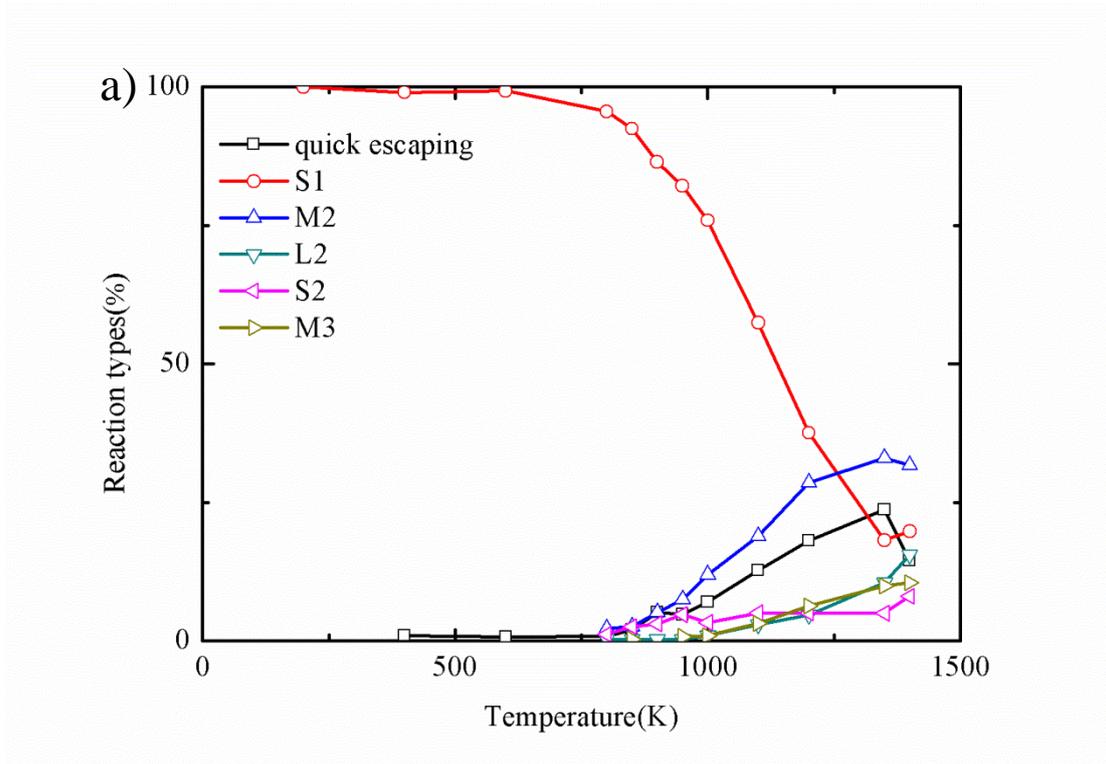

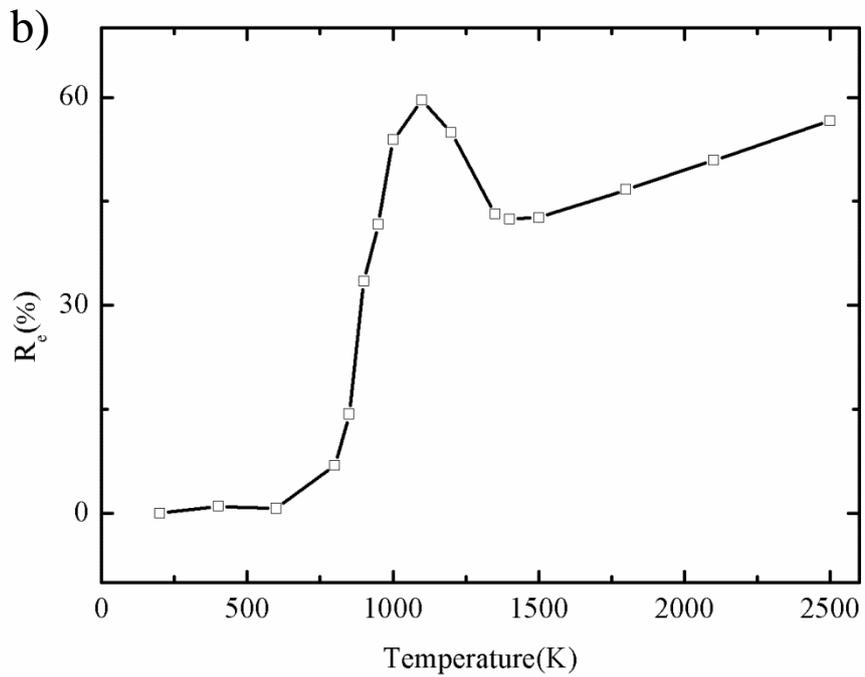

**Fig. 6.** (a) The percentage of different types of trap mutations and quick-escaping He atoms of the total absorbed He atoms as a function of temperature; (b) The percentage of He atoms escaping into vacuum of all the absorbed He atoms.

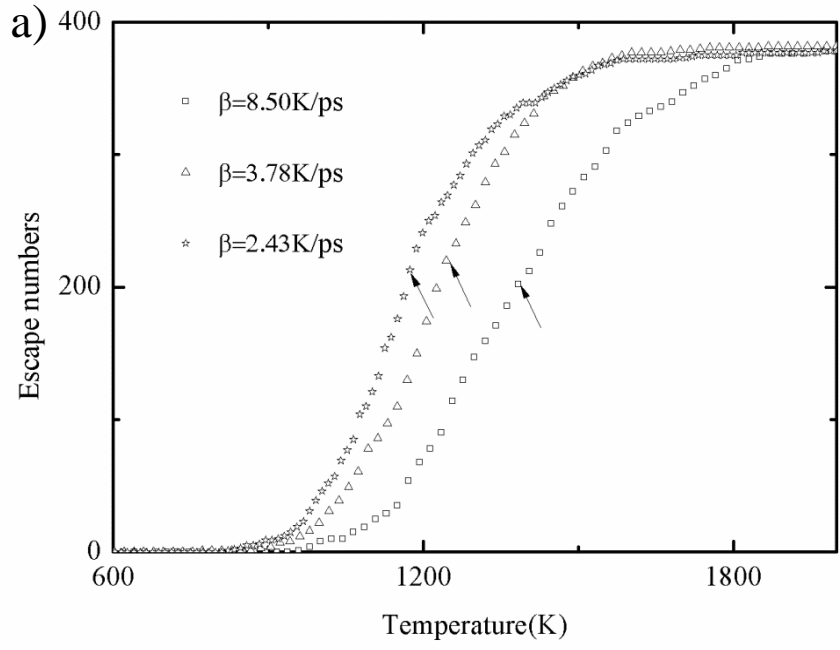

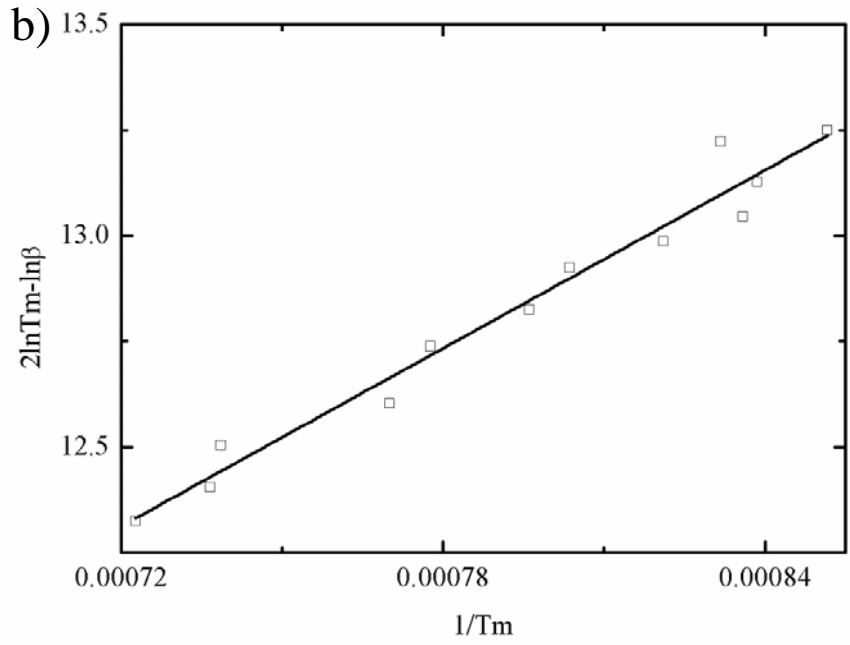

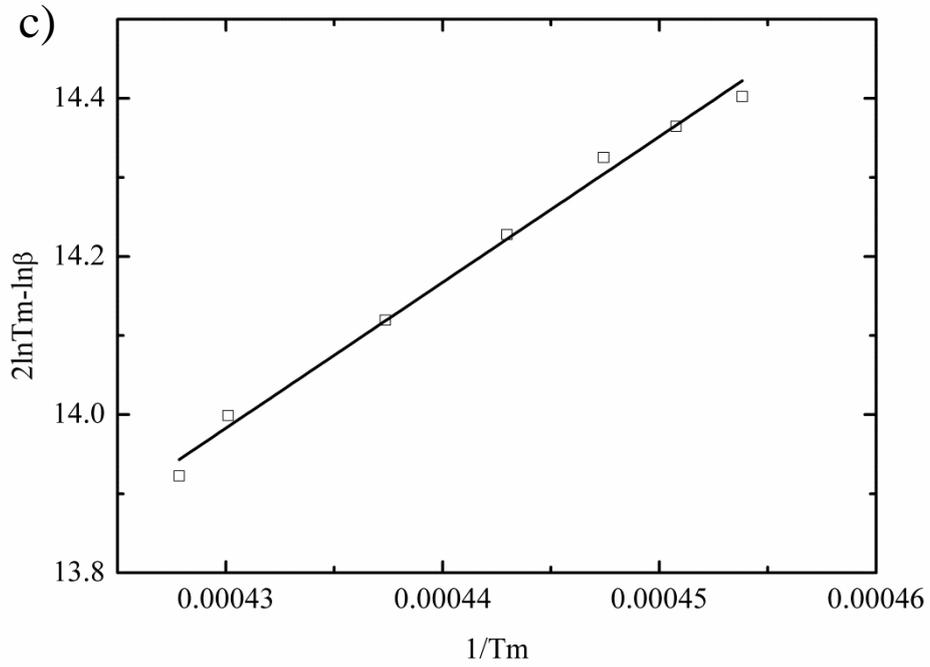

**Fig. 7.** (a) The accumulated number of He atoms escaping into vacuum from $S_1$ trap mutations in MD simulations of thermal desorption. The arrows point to the $T_m$.; (b) $2\ln T_m - \ln\beta$ vs. $1/T_m$ for the $S_1$ trap mutations and (c) for the $M_2$ trap mutations. Symbols: the data of MD simulations; Solid line: eq. (3) fitted to the MD data (refer to eq. (3) in the text).

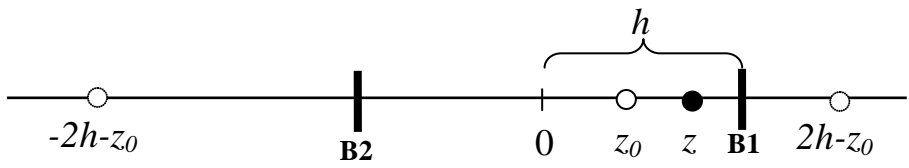

**Fig. A.1**. A schematic diagram explaining the concept of virtual source.